\documentclass[preprint,showpacs,preprintnumbers,amsmath,amssymb,natbib]{revtex4}

\usepackage{graphicx}
\usepackage{epsfig}		
\usepackage{dcolumn}
\usepackage{bm}
\def\0{\mbox{\tiny $0$}}
\def\1{\mbox{\tiny $1$}}
\def\2{\mbox{\tiny $2$}}
\def\3{\mbox{\tiny $3$}}
\def\4{\mbox{\tiny $4$}}
\def\5{\mbox{\tiny $5$}}
\def\6{\mbox{\tiny $6$}}
\def\7{\mbox{\tiny $7$}}
\def\8{\mbox{\tiny $8$}}
\def\9{\mbox{\tiny $9$}}

\def\f14{\mbox{\tiny $\frac{1}{4}$}}

\begin{document}

\title{Topological origin of quantum mechanical vacuum transitions and tunneling}

\author{Alex E. Bernardini}
\email{alexeb@ufscar.br}
\author{Mariana Chinaglia}
\email{chinaglia@ufscar.br}
\affiliation{Departamento de F\'{\i}sica, Universidade Federal de S\~ao Carlos, PO Box 676, 13565-905, S\~ao Carlos, SP, Brasil}

\date{\today}

\begin{abstract}
The quantum transition between shifted zero-mode wave functions is shown to be induced by the systematic deformation of topological and non-topological defects that support the $1$-dim double-well (DW) potential tunneling dynamics.
The topological profile of the zero-mode ground state, $\psi_{0}$, and the first excited state, $\psi_{1}$, of DW potentials are obtained through the analytical technique of topological defect deformation.
Deformed defects create two inequivalent topological scenarios connected by a symmetry breaking that support the quantum conversion of a zero-mode stable vacuum into an unstable tachyonic quantum state.
Our theoretical findings reveal the topological origin of two-level models where a non-stationary quantum state of unitary evolution, $\psi_{0} + e^{\mbox{\tiny$-i E \,t$}}\psi_{1}$, that exhibits a stable tunneling dynamics, is converted into a quantum superposition involving a self-vanishing tachyonic mode, $e^{\mbox{\tiny$- E \,t$}}\psi_{0} + \psi_{1}$, that parameterizes a tunneling coherent destruction.
The non-classical nature of the symmetry breaking dynamics is recreated in terms of the single particle quantum mechanics of $1$-dim DW potentials.
\end{abstract}


\date{\today}
\maketitle

\section{Introduction.}

The topological origin of quantum phase-transitions has been conjectured in many different frameworks in theoretical and experimental physics.
In the phenomenological context, for instance, topological phase transitions have been considered in identifying the equivalence between the Harper and Fibonacci quasicrystals \cite{Verbin,Kraus}, in examining Golden string-net models with Fibonacci anyons \cite{Schulz}, in quantifying the stability of topological order of spin systems perturbed by magnetic fields \cite{Trebst}, and even for describing phase transitions in models of DNA \cite{DNA}.
Likewise, theoretical tools have been developed for describing mean-field XY models which are related to suitable changes in the topology of their configuration space \cite{Casetti}, for investigating the relation between the strong correlation and the spin-orbit coupling for Dirac fermions \cite{Yu}, and for conjecturing that second-order phase transitions also have topological origin \cite{Caiani}.
Also in quantum field theories \cite{Cvetic01}, tachyonic modes \cite{Sen,Campos,Bertolami} can be realized by the instability of the quantum vacuum, described by the quantum state displaced from a local maximum of an effective topological potential, $ U(\xi)$.
It implies into a process called tachyon condensation \cite{PRL1,PRL2}, which also exhibits a topological classification. 

Notwithstanding the ferment in such an enlarged scenario, the topological classification might be relevant because it encodes the information about distortions and deformations of quantum systems. 
A quantum phase transition is indeed supposed to occur when a system can be continuously deformed into another system with distinct topological origins.
Our work is therefore concerned with obtaining the topological origin of quantum states, and the corresponding quantum mechanical (QM) potentials, that describe the scenario of single-particle tunneling in double-well (DW) potentials \cite{caticha}. 

The coherent control of single-particle tunneling in strongly driven DW potentials have been recurrently investigated as they appear in several relevant scenarios \cite{Schulz,Domcke}.
Analytically solvable DW models \cite{Keung} are also supposed to be prominent candidates for benchmarks of computer codes and numerical networks.
Through a systematic topological defect deformation \cite{Bas01,AlexRoldao}, here it is shown that the zero-mode stable vacuum (ground state) supported by a topological defect can be converted into an unstable (tachyonic) quantum state supported by the corresponding deformed topological defect.
The topological (vacuum) symmetry breaking driven by a phenomenological parameter, $\sigma$, converts a stable (unitary and non-stationary) quantum tunneling configuration, $\psi_{0} + e^{\mbox{\tiny$-i t/\sigma$}}\psi_{1}$, into an unstable (non-unitary) quantum superposition involving a self-vanishing tachyonic mode, $e^{\mbox{\tiny$- t/\sigma$}}\psi_{0} + \psi_{1}$.
A remarkable and distinctive feature is that even with vacuum symmetry breaking, the topological origin is recovered.
As a residual effect, $\psi_{1}$ becomes the novel zero-mode stable solution supported by the deformed defect.
Our results reveal that the tunneling dynamics can be brought to stationary configurations that mimics the complete standstill known as coherent destruction of tunneling \cite{Grossmann,Kierig}.
The evolution of both non-stationary and non-unitary quantum perturbations supported by non-equivalent topological profiles are described through the Wigner's quasi-probability distribution function \cite{Ritter}.

\section{Topological scenarios for DW potentials}

Let us consider the simplest family of $1+1$-dim classical relativistic field theories for a scalar field, $\varphi(t,s)$, described by a Lagrangian density that leads to the equation of motion given by
\begin{equation}
\partial_t^{\2} \varphi - \partial_s^{\2} \varphi + dU/d\varphi = 0,
\label{KG}
\end{equation}
from which, for the most of the known theories driven by $V(\varphi)$, the scalar field supports time-independent solutions, $\xi(s)$, of finite energy \cite{Rajaraman,Coleman}.
The field solution stability can be discussed in terms of small time-dependent perturbations given by
$\delta(t,s) \sim \varphi(t,s) - \xi(s)$. 
Rewriting Eq.~(\ref{KG}) in terms of $\xi$ and $\delta$ and retaining only terms of first-order, one finds
$\square^{\2} \delta + (d^2U/d\xi^2)\delta = 0$,
from which the time translation invariance implies that $\delta$ can be expressed as a superposition of normal quantum modes as
$\delta(t,s) = \sum_{n} a_{n} \, \exp{(i\,\omega_n\, t)} \psi_{n}(s)$, where $a_{j}$ are arbitrary coefficients and $\psi_j$ and $\omega_j$ obey the $1$-dim Schr\"odinger-like equation,
\begin{equation}
\left(- d^2/ds^2 + V_0^{(QM)}(s) \right)\psi_{n}(s)= \omega_n^2\,\psi_{n}(s),
\label{SC}
\end{equation}
and where the quantum mechanical (QM) potential is identified by $V_0^{(QM)}(s) = d^2U(s)/d\xi^2(s)$.
Stationary modes require that $\omega_n^2 > 0$.
Likewise, unstable tachyonic modes can be found if $\omega_n \sim i \kappa_n$, with $ - \omega_n^2 = \kappa_n^2 > 0$.

For scalar field potentials, $U(\xi)$, that engender kink- and lump-like structures \cite{Bas01,Bas02,Bas03}, the first-order framework \cite{BPS,Rajaraman,Bas01} sets
$U(\xi) = z_{\xi}^{2}/2,~\mbox{with}~z_{\xi} = dz/d\xi = d\xi/ds = \xi^{\prime}$,
and Eq.~(\ref{SC}) can be written as
\begin{equation}
\left(- d^2/ds^2 + \xi^{\prime\prime\prime}/\xi^{\prime}\right)\psi_{n} = \omega_n^2\,\psi_{n}.
\label{SC2}
\end{equation}
A simple mathematical manipulation allows one to obtain the zero-mode, $\psi_{0}$ (i. e. when $\omega_0 = 0$), as $\psi_{0} \propto \xi^{\prime}$, which might correspond to the quantum ground state in case of $\xi^{\prime}$ describing lump-like functions having no zeros (nodes).
In this case, there would be no instability from modes with $\omega_n^2 < 0$, and hence building a suitable analytical model for topological scenarios turns into a simple matter of finding integrable solutions.

Given Eq.~(\ref{SC2}), the ground and first excited states of some few 1-dim quantum systems supported by (at least partially) exactly solvable DW potentials can be expressed in analytical closed forms \cite{caticha} through a multiplier function constraint, $\alpha$, defined by $\psi_1(s) = \alpha(s) \psi_0(s)$.
It is analogous to super-symmetric QM procedures for computing the ground state and the corresponding potential \cite{gend,cooper}.
Substituting $\psi_1(s) = \alpha(s) \psi_0(s)$ into Eq.~(\ref{SC}), with $n=1$, one has 
\begin{equation}
(\alpha(s) \psi_0(s))^{\prime\prime} +(\omega^{2}_1 - V_0^{(QM)}(s))\alpha(s) \psi_0(s)= 0,
\end{equation}
and subtracting Eq.~(\ref{SC}) with $n=0$,
\begin{equation}
\psi_0^{\prime\prime}(s) +(\omega^{2}_0 - V_0^{(QM)}(s))\psi_0(s)= 0,
\end{equation}
one obtains
\begin{equation}
\alpha^{\prime\prime} (s) + 2 \alpha^{\prime}(s)\beta(s)+\Delta\omega^{2}_{10} \alpha(s)= 0,
\end{equation}
with $\beta(s) = \ln(\psi_0(s))^{\prime}$, and $\Delta\omega^{2}_{10} = \omega^{2}_1 - \omega^{2}_0$.
The corresponding ground state is therefore
$\psi_0(s) = \mathcal{N} \,\exp\left(\int^s ds^{\prime} \, \beta(s^{\prime})\right)$,
where $\mathcal{N}$ is a normalization constant, and the DW potential can be computed through
\begin{equation}
V_0^{(QM)}(s) = (\psi^{\prime\prime}_0(s)/\psi_0(s)) + \omega^{2}_0,
\end{equation}
which can be simplified for zero-modes with $\omega^{2}_0 = 0$.

A set of symmetric and asymmetric DW potentials satisfying the above properties can be identified when $\alpha(s)$ presents an asymptotic behavior of kink-like functions, for instance, when
\begin{equation}
\alpha(s) = \epsilon + \tanh(\gamma s),
\label{relation}
\end{equation}
where the asymmetry is determined by $\epsilon \neq 0$, and $\gamma$ is an arbitrary constant.
From Eq.(\ref{relation}) one can realize that the parameter $\epsilon$ determines the asymmetry, seen as a small deformation of the initial system. The proposal of deformation of topological defects has been introduced to 
 drive perturbations of original topological defects and, in particular, to generate the different topological sectors of double and triple sine-Gordon theories, as well as subsequent applications in asymmetric brane-world models \cite{Baz14,Baz14B}.

\begin{figure}
\includegraphics[scale=0.56]{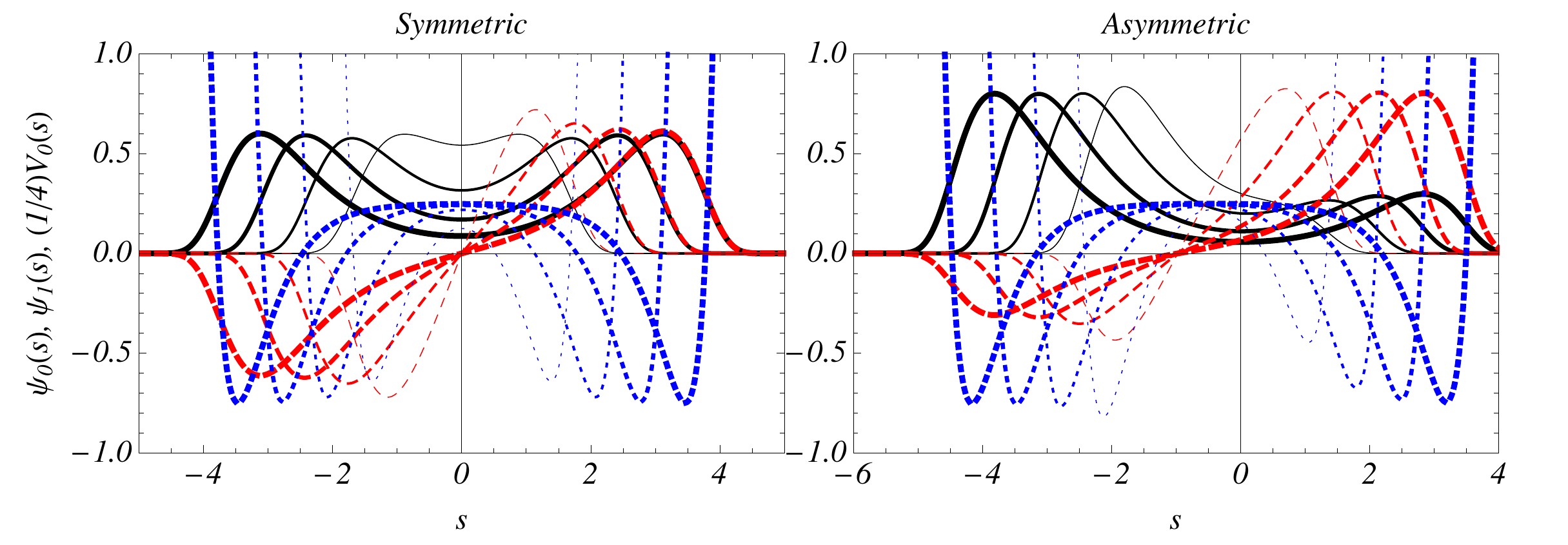}
\vspace{-.7 cm}
\caption{
\small (Color online) Symmetric (first plot, $\epsilon = 0$) and asymmetric (second plot, $\epsilon = 3/4$)  quantum DW  potentials (dotted blue lines), $V_0^{(QM)}$, the corresponding normalized ground state (solid black lines), $\psi_0$, and first excited state (dashed red lines), $\psi_0$.
The plots are for $\gamma = 1$ and $\omega_1 = 1/\sigma$, with the parameter $\sigma = 1$ (thinest line), $2,\,4$ and $8$ (thickest line).}
\label{Topo01}
\vspace{-.6 cm}
\end{figure}

Assuming a zero-mode ($n=0$) state with $\omega_0 = 0$, the ground and first excited states are obtained as
\small
\begin{eqnarray}
\psi_{0}(s) \hspace{-.15cm}&=&\hspace{-.15cm}\mathcal{N}\cosh(\gamma s)\exp\left[-\frac{\cosh(2\gamma s) + \epsilon(2\gamma s +\sinh(2\gamma s) )}{8\gamma^2 \sigma^2}\right],\nonumber\\
\psi_{1}(s) \hspace{-.15cm}&=& \hspace{-.15cm}\mathcal{N}\sinh(\gamma s)\exp\left[-\frac{\cosh(2\gamma s) + \epsilon(2\gamma s +\sinh(2\gamma s) )}{8\gamma^2 \sigma^2}\right],
\label{eigenfunctions}
\end{eqnarray}
\normalsize
where we have assumed that $\omega_1 = 1/\sigma$.
The above normalized stationary states, as well as the corresponding DW potential, can be depicted in Fig.~\ref{Topo01}, for $\epsilon = 0$ and $3/4$.
By matter of convenience, the following discussion shall be concerned only with the symmetric case ($\epsilon = 0$). Straightforward extensions to asymmetric potentials ($\epsilon \neq 0$) are shown in the end.

\begin{figure}\hspace{-.2cm}
\includegraphics[scale=0.56]{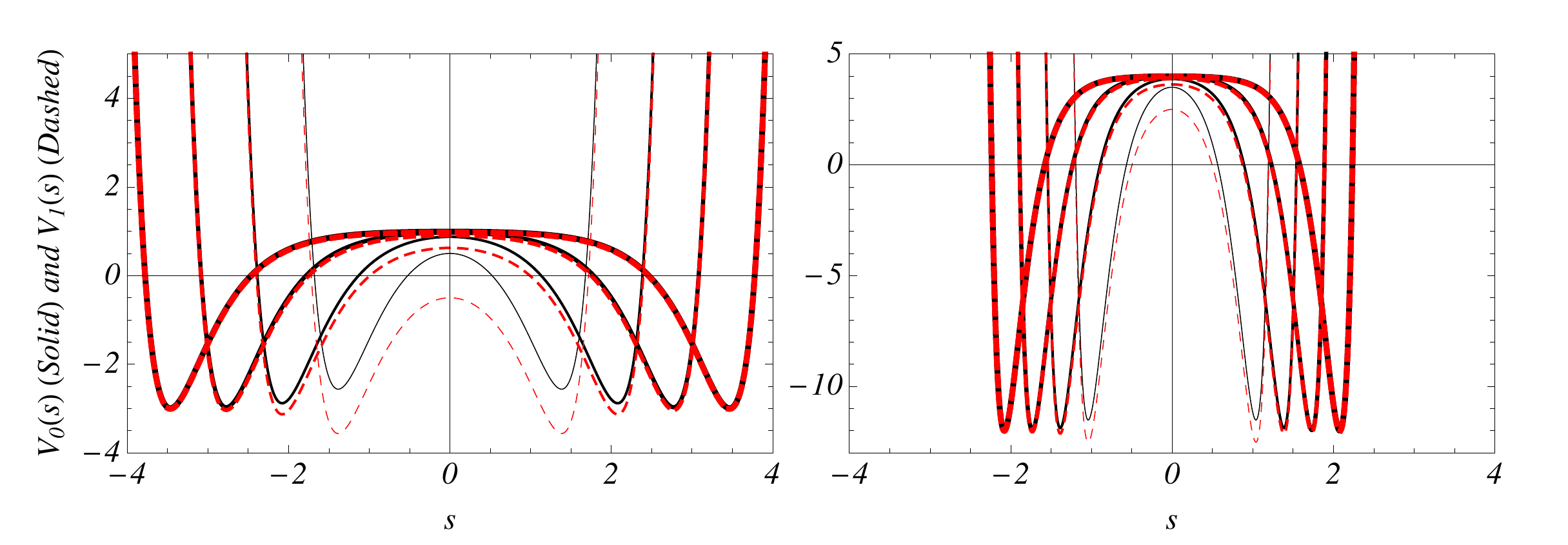}
\vspace{-.7 cm}
\caption{
\small (Color online) Quantum potentials, $V^{(QM)}_{0}(s)$ (solid black lines) and $V^{(QM)}_{1}(s)$ (dashed red lines), supported by topological scenarios driven by scalar fields, $\xi$ and $\chi$, respectively.
The first and second plots are for $\gamma = 1$ and $2$, respectively, and again, $\omega_1$ was set equal to $1/\sigma$, with the parameter $\sigma = 1$ (thinest line), $2,\,4$ and $8$(thickest line).}
\label{Topo02}
\end{figure}

As previously explained, knowing the zero-mode, $\psi_{0}$, allows for identifying the corresponding topological scenario that supports quantum perturbations for the above DW potentials, $V^{(QM)}_{0}(s)$.
For a potential $U(\xi)$ driven by a scalar field $\xi$, with $V^{(QM)}_{0}(s) = \xi^{\prime\prime\prime}/\xi^{\prime}$, it can be identified as $\psi_{0} \propto  \xi^{\prime}$.
Moreover, one should notice that a deformed DW model can be identified by shifting the energy of the first excited state from $\omega_1$ to $0$.
In this case, one should infer the existence of a novel quantum potential, $V^{(QM)}_1(s)$, such that
\begin{equation}
\left(- d^2/ds^2 + V^{(QM)}_1(s)\right)\psi_{1} = 0,
\end{equation}
and therefore $\psi_1 = \alpha \psi_0$ implies that
\begin{equation}
\left(- d^2/ds^2 + V^{(QM)}_1(s)\right)\psi_{0} = -\omega_1^2\,\psi_{0} =\kappa_1^2\,\psi_{0}.
\end{equation}
The eigenfunction $\psi_1(s)$ is then identified as the novel zero-mode and
the deformed QM potential thus supports an unstable tachyonic mode, $\psi_{-1}(s,t)$, that evolves in time as a decaying state,
\begin{equation}
\psi_{-1}(t,s) \equiv \exp{(-\omega_{1} t)}\psi_{0}(s).
\end{equation}
Again, the novel zero-mode, $\psi_{1}$, allows for identifying the corresponding topological scenario for $V^{(QM)}_{1}(s)$.
In Fig.~\ref{Topo02}, one can depict the QM potentials,
\small
\begin{equation}
V^{(QM)}_{0(1)}(s) =\frac{16\gamma^2\sigma^2\left[(-1)^{0(1)} + 2 ( \gamma^2\sigma^2- \cosh(2\gamma s))\right] + \cosh(4\gamma s) }{32\gamma^2\sigma^4},
\end{equation}\normalsize
which support the zero-modes: the primitive one, $\psi_0$, and the deformed one, $\psi_1$, respectively (c. f. Eq.~(\ref{eigenfunctions})).  

In the asymptotic limit of $\sigma^2 \rightarrow \infty$, which corresponds to $\omega_1\rightarrow 0$ , the potential tends to an asymptotic constant value $\sim \gamma^{\2}$, along which the quoted continuous transition between stable and unstable regimes occurs.
To illustrate such a continuous deformation, we define a phenomenological parameter $\epsilon = 1/\sigma$ such that the potentials $V^{(QM)}_{0}$ and $V^{(QM)}_{1}$ can be generically written as 
\begin{equation}
V^{(QM)}(s) =\frac{16\gamma^2\left[\epsilon\vert\epsilon\vert + 2 ( \gamma^2- \epsilon^2\cosh(2\gamma s))\right] + \epsilon^4\cosh(4\gamma s) }{32\gamma^2},
\end{equation}\normalsize
The deformation destroys the potential profile at the transition point, $\epsilon = 0 $, for which $V^{(QM)}(s) = \gamma^{\2}$.
We shall show in the next section
that an exact deformed framework for a novel potential $W(\chi)$ driven $\chi$, with $V^{(QM)}_{1}(s) = \chi^{\prime\prime\prime}/\chi^{\prime}$, can be obtained if one identifies $\psi_{1}$ as  $\propto  \chi^{\prime}$.
The question to be posed 
is mainly concerned with the viability of obtaining and connecting topological scenarios that support both stable and tachyonic eigenfunctions.

\section{Stable and tachyonic modes supported by deformed defects and related quantum transitions}

Suitable techniques have been suggested to study and solve non-linear equations through deformation procedures.
In particular, obtaining static structures of localized energy through deformed defects allows for modifying a primitive defect structure - for instance, the kink defect from the $\lambda \phi^4$ theory - by means of successive deformation operations.

Let us then consider a sequence of defect deformation involving two different topological scenarios, namely $\xi(s)$ and $\chi(s)$, with $s$ as the spatial coordinate, in a way that the deformation procedure is prescribed by the following first-order equations \cite{Bas01,Bas02,Bas03,AlexRoldao},
\begin{eqnarray}
\xi^{\prime} \equiv \frac{d \xi}{d s} = z_{\xi} &=&  \xi_{\phi} y_{\phi},\nonumber\\
\chi^{\prime} \equiv \frac{d \chi}{d s} = w_{\chi} &=&  \chi_{\phi} y_{\phi} = \chi_{\xi} z_{\xi},
\label{topo02}
\end{eqnarray}
where $\xi_{\phi}$, $\chi_{\phi}$, and $\chi_{\xi}$ are invertible deformation functions such that $\xi_{\phi} \chi_{\xi} = \chi_{\phi}$, with subindices that stand for the corresponding derivatives. 
The BPS framework \cite{BPS,Bas01,Rajaraman} states that derivatives of auxiliary superpotentials, $z_{\xi}$, $w_{\chi}$, and $y_{\phi}$, through Eqs.~(\ref{topo02}) can be used to build a deformation chain as
\begin{equation}
V(\phi) = \frac{1}{2} y_{\phi}^2 ~\rightleftarrows ~
W(\chi) = \frac{1}{2} w_{\chi}^2 ~\rightleftarrows ~
U(\xi) = \frac{1}{2} z_{\xi}^2.
\label{topo02B}
\end{equation}

Turning back to our QM problem, one can assume that $\alpha(s)$ (c. f. Eq.~(\ref{relation})) corresponds to a deformation function, 
\begin{equation}
\alpha(\phi) = \chi_{\xi} = \frac{\chi_{\phi}(\phi)}{\xi_{\phi}(\phi)}.
\label{alfa}
\end{equation}
Thus, according to Eq.~(\ref{topo02}), the following deformation chain can be established,
\begin{equation}
\psi_{1} \propto  \chi^{\prime} \equiv w_{\chi} = \chi_{\xi} z_{\xi} = \alpha z_{\xi} \equiv \alpha \xi^{\prime} \propto \alpha \psi_{0},
\end{equation}
and one obtains analytical expressions for $z_{\xi}$ and $w_{\chi}$ as functions of $\phi\equiv\phi(s)$.
Once the primitive defect, $\phi$, is identified as the kink of the $\lambda \phi^4$ theory, i. e. $\phi(s) = \pm \tanh(s)$ and $y_{\phi} = \phi^{\prime} = (1-\phi^{\2})$, one obtains two first-order differential equations,
\begin{equation}
\xi_{\phi}(\phi) =  z_{\xi}(\phi)/y_{\phi}(\phi) ~~\mbox{and}~~
\chi_{\phi}(\phi) =  w_{\chi}(\phi)/y_{\phi}(\phi),
\label{phiphi}
\end{equation}
which are obviously constrained by $\alpha(\phi) = \chi_{\xi}$.
Finally, the analytical results for $(\xi, z^{2}_{\xi}/2)$ and  $(\chi, w^{2}_{\chi}/2)$ as functions of $\phi$ allows one to straightforwardly obtain an analytical parametric representation for $U(\xi)$ and $W(\chi)$, respectively, the topological potentials that support the quantum dynamics driven by $V^{(QM)}_0$ and $V^{(QM)}_1$.
Given the topology of the deformation function, $\alpha(\phi)$, and the correspondence with Eq.~(\ref{topo02B}), one notices that kinks are deformed into lumps and vice-versa.

\begin{figure}
\includegraphics[scale=0.56]{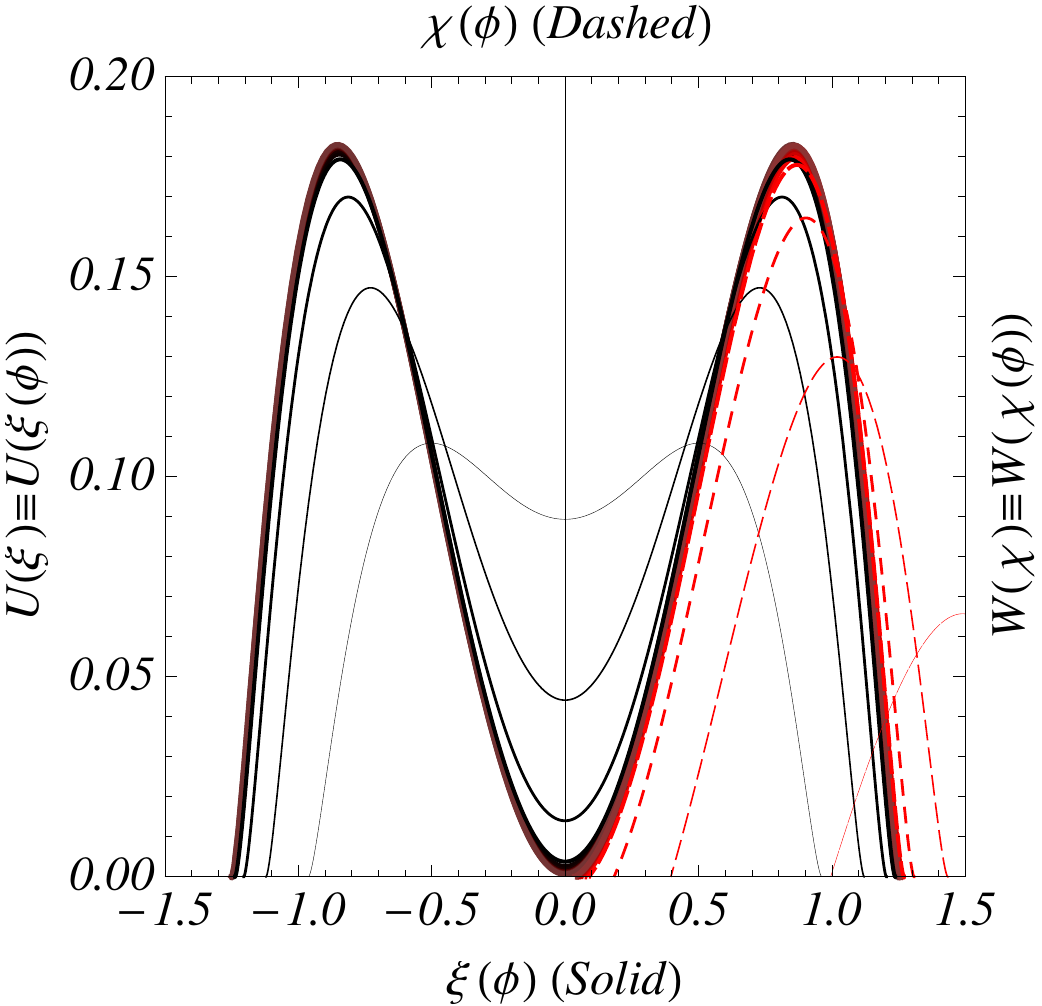}
\hspace{.5 cm}
\includegraphics[scale=0.56]{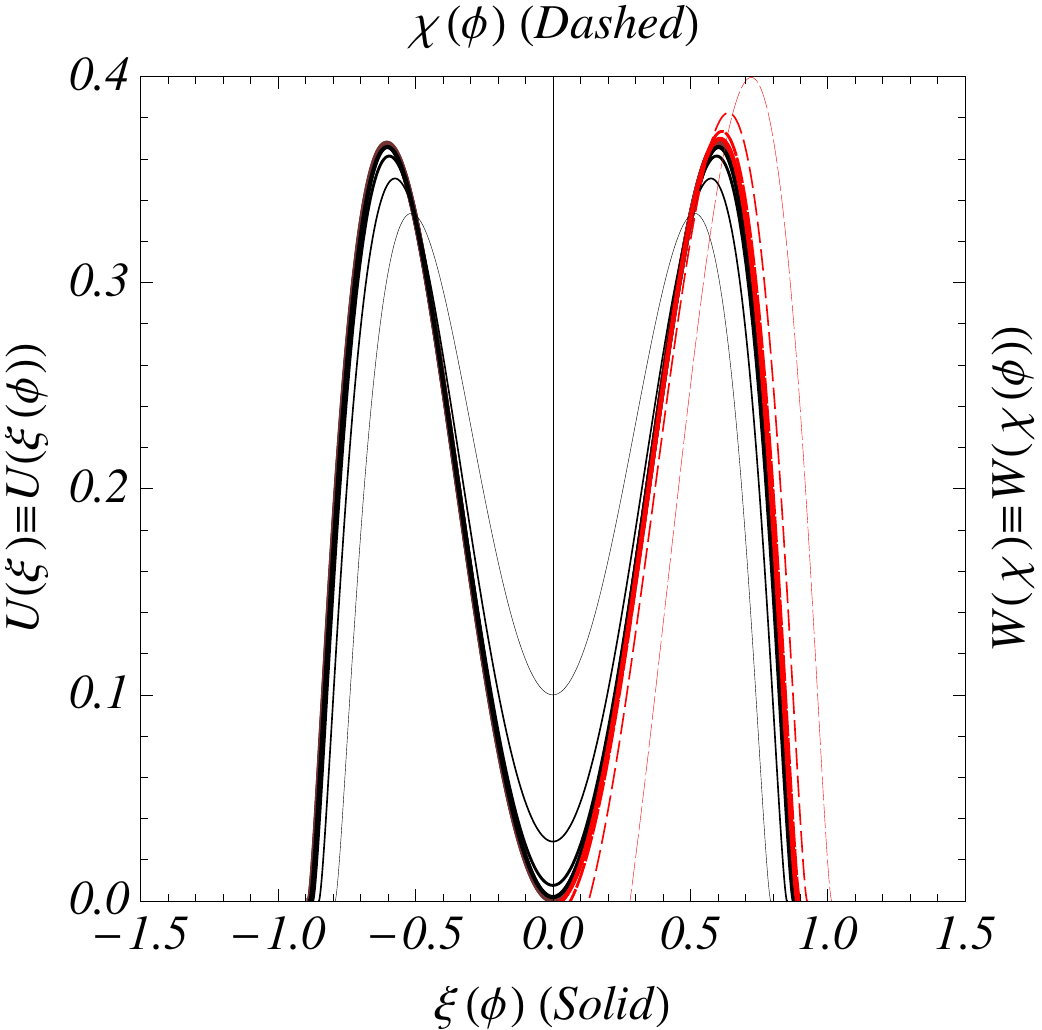}
\includegraphics[scale=0.56]{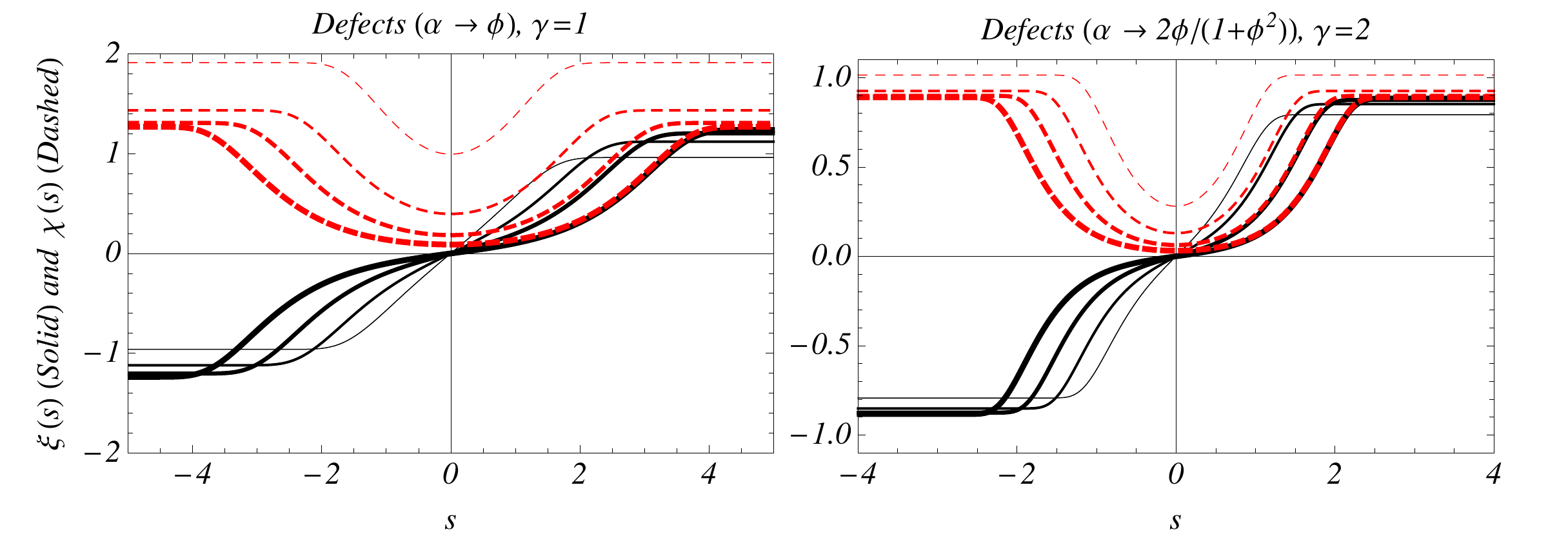}
\vspace{-.7 cm}
\caption{
\small (Color online) (First row) Potentials, $U(\xi)$ (solid black lines) and $W(\chi)$ (dashed red lines), as a parametric function of $\phi$. The black to red (grayscale) transition region illustrates the symmetry breaking that sets the continuous transition from parameters $\omega_1^2 > 0$ to $\omega_1^2 = -\kappa_1^2 < 0$.
(Second row) Correspondence with deformed defects, $\xi(s)$ and $\chi(s)$.
The plots are for $\gamma = 1$ (first column) and $2$ (second column) where $\omega_1 = 1/\sigma $ (solid black lines) and $\kappa_1 = 1/\sigma$ (dashed red lines) with $\sigma = 1$ (thinest line), $2,\,4$ and $8$ (thickest line).}
\label{Topo04}
\vspace{-.6 cm}
\end{figure}

To clear up this point, let us consider two particular cases of the above QM symmetric DW potential theory by attributing discrete values to $\gamma$ into Eq.~(\ref{relation}).
For $\gamma =1$, and following the convention that reduces our analysis to $+$ sign solutions for $\phi(s)$, one has $\alpha(\phi) = \phi$, that gives
\begin{equation}
z_{\xi}(\phi) =
\frac{e^{\left(-\frac{3-2\phi^2}{8 \sigma^2 (1-\phi^2)}\right)}}{\sqrt{1-\phi^2}}
~~\mbox{and}~~
w_{\chi}(\phi)= \phi\,z_{\xi}(\phi),
\end{equation}
and, after integrating Eqs.~(\ref{phiphi}),
\begin{eqnarray}
\xi(\phi) &=& 
\frac{\sqrt{\pi}\sigma}{ e^{\left(1/(8\sigma^2)\right)}}
\mbox{Erf}\left[\frac{1}{2\sigma}\frac{\phi}{\sqrt{1-\phi^2}}\right]~~ \mbox{and}
\nonumber\\\chi(\phi)&=&
\sqrt{\pi} \sigma e^{\left(1/(8\sigma^2)\right)}
\mbox{Erf}\left[\frac{1}{2\sigma}\frac{1}{\sqrt{1-\phi^2}}\right],
\end{eqnarray}
where we have suppressed the normalization constant, $\mathcal{N}$, from the notation.
For $\gamma = 2$, one has $\alpha(\phi) = 2\phi/(1 + \phi^2)$ that gives
\begin{equation}
z_{\xi}(\phi) = 
\frac{1+\phi^2}{1-\phi^2}
e^{\left(-\frac{1+2\phi^2+\phi^4}{16\sigma^2(1+\phi^2)^2}\right)}
~~\mbox{and}~~
w_{\chi}(\phi)= 
\frac{2\phi}{1+\phi^2}
z_{\xi}(\phi)
\end{equation}
and, after integrating Eqs.~(\ref{phiphi}),
\begin{eqnarray}
\xi(\phi) &=& 
\frac{\sqrt{\pi}\sigma}{ e^{\left(1/(32\sigma^2)\right)}}
\mbox{Erf}\left[\frac{1}{2\sigma}\frac{\phi}{1-\phi^2}
\right]~~ \mbox{and}\nonumber\\
\chi(\phi)&=&
\sqrt{\pi}\sigma e^{\left(1/(32\sigma^2)\right)}
\mbox{Erf}\left[\frac{1}{4\sigma}\frac{1+\phi^2}{1-\phi^2}
\right].
\end{eqnarray}
The BPS potentials, $U(\xi)$ and $W(\chi)$, for both examples can be depicted in the first row of Fig.~\ref{Topo04}.
The corresponding topological defects, $\xi(s)$ and $\chi(s)$, are depicted in the second row of Fig.~\ref{Topo04}.
The kink-like structures, $\xi(s)$, have the topological index given by $Q(\sigma) = 2 f_{\gamma(\sigma)}$, for $\gamma = 1,\,2$. For lump-like structures, $\chi(s)$, the topological index vanishes.
Turning the energy eigenvalue parameter into a phenomenological variable, one can say that the situation described in Fig.~\ref{Topo04} resembles conventional phase transitions accompanied by spontaneous symmetry breaking, as those that results into the formation of topological defects via the Kibble-Zurek mechanism \cite{Kibble01}.
With the phenomenological parameter $\omega_1^2 = \sigma^{-2}$ running from positive to negative values, through a continuous ($1/\sigma$) transition, the topological symmetry breaking converts a non-stationary quantum state of unitary (stable) evolution, $\Psi_S \sim \psi_{0} + e^{\mbox{\tiny$(-i t/\sigma)$}}\psi_{1}$, into a non-unitary (unstable) quantum superposition involving a suppressing tachyonic mode, $\Psi_U \sim e^{\mbox{\tiny$(- t/\sigma)$}}\psi_{0} + \psi_{1}$. 
The tunneling pattern is destroyed since $\Psi_{U}$ is collapsed into the coherent (zero-mode) stable solution, $\psi_1$, supported by the deformed defect.

\begin{figure}
\includegraphics[scale=0.47]{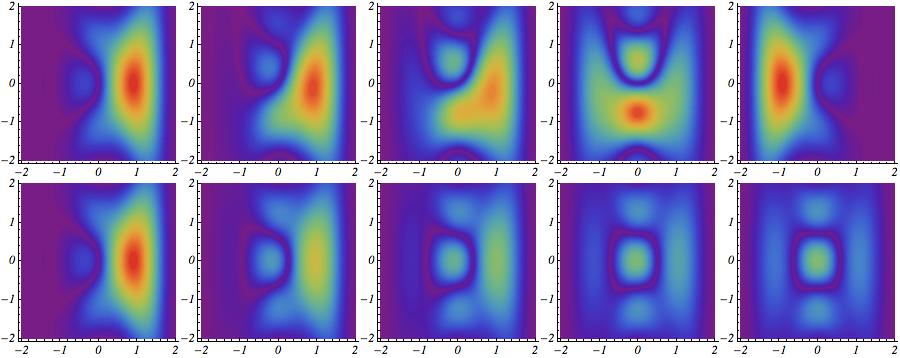}
\caption{
\small (Color online) Stable (non-stationary) and unstable (non-unitary) symmetric tunneling evolution described by the Wigner functions (in modulus) for $\Psi_S(W_S)$ (first row) and $\Psi_U(W_U)$ (second row), respectively .
The plots are for times $T = 0$ (left), $\pi/8,\,\pi/4,\,\pi/2$, and $\pi$ (right) and $\gamma = \sigma = 1$.
The contour plot follows a rainbow color scheme from red (light gray) which corresponds to $1$, to violet (dark gray) which corresponds to $0$.
At time $T=0$ the Wigner functions, $W_{S(U)}(s,p;T)$ are assumed to be centered at the origin of phase-space coordinates, $(0, 0)$.}
\label{Wig02}
\end{figure}

\begin{figure}\hspace{-.2cm}
\includegraphics[scale=0.56]{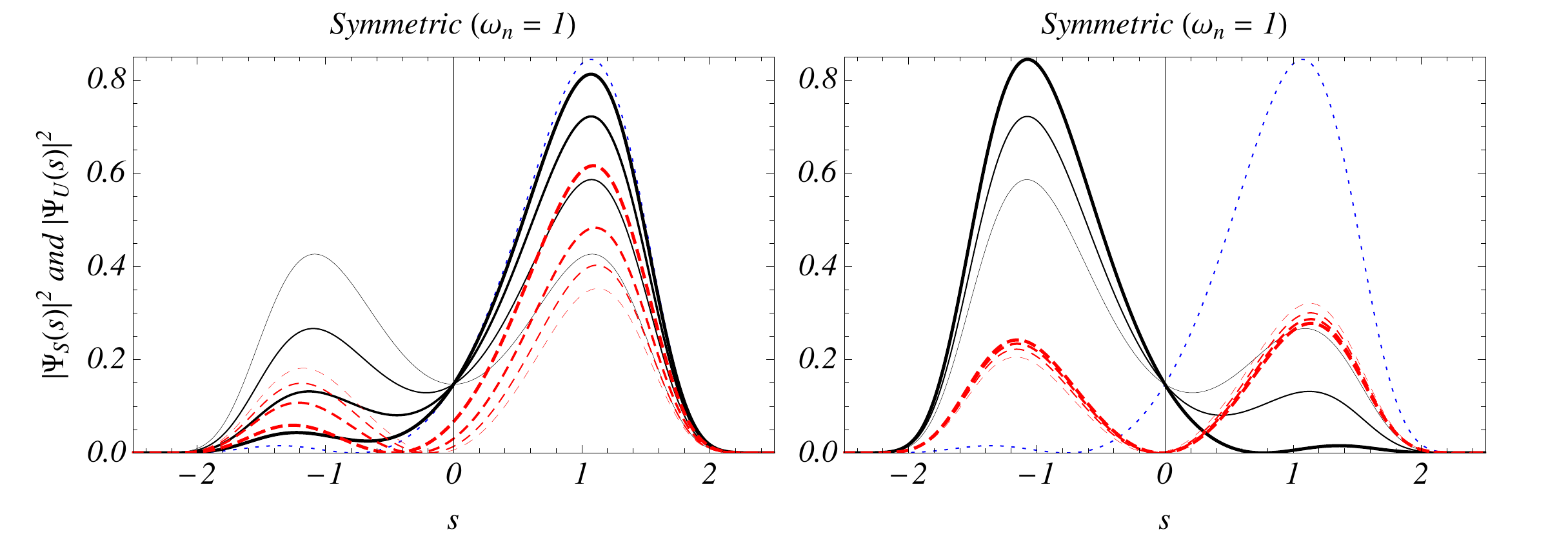}
\caption{
\small (Color online) Time evolution of probability densities for stable (solid black lines) and unstable (dashed red lines) composite states, $\Psi_S$ and $\Psi_U$, in a DW symmetric configuration.
The curves are for times running from $T = \pi/8$ (thickest line) to $T = \pi/2$ (thinest line) in the first plot, and from $T = 5\pi/8$ (thinest line) to $T = \pi$ (thickest line) in the second plot, with steps $\Delta T = \pi/8$.
Dotted blue lines correspond to $T = 0$ for both, stable and unstable tunneling.
The results are for $\gamma = \sigma = 1$.}
\label{Wig01}
\end{figure}

To investigate the tunneling dynamics of both stable, $\Psi_S$, and unstable, $\Psi_U$, configurations,
the phase-space representation of quantum dynamics may be a kind of elucidative in identifying quantum transitions, as one can notice from the Wigner quasi-probability distribution depicted in Fig.~\ref{Wig02} as given by
\begin{equation}
W_{S(U)}^{(s,p;t)} = (\pi\hbar)^{-1}\int_{-\infty}^{+\infty}\hspace{-.3cm} dy\,
\Psi_{S(U)}^{*(s+y;t)}\Psi_{S(U)}^{(s-y;t)}\,e^{\frac{2i\,p\,y}{\hbar}}.
\end{equation}
The corresponding time evolution of probability densities for $\Psi_S$ and $\Psi_U$ are also shown in Fig.~\ref{Wig01}.
One infers that the smaller is the energy splitting parameter, $\sigma$, the larger is the beat period for the stable configurations and the decaying time for the unstable ones.
Extending the Wigner map along the axis of momentum coordinate for a single DW converted into a periodic chain allows one to reproduce the results of a driven DW realized by quantum configurations of periodically curved optical wave guides \cite{Valle} where the spatial light propagation imitates the space-time dynamics of matter waves in a DW governed by the Schr\"odinger equation.
The same interpretation is valid for periodic potential quantum systems that support the nonlinear dynamics of a Bose-Einstein condensates \cite{Lignier}.

\begin{figure}[h!]
\includegraphics[scale=0.56]{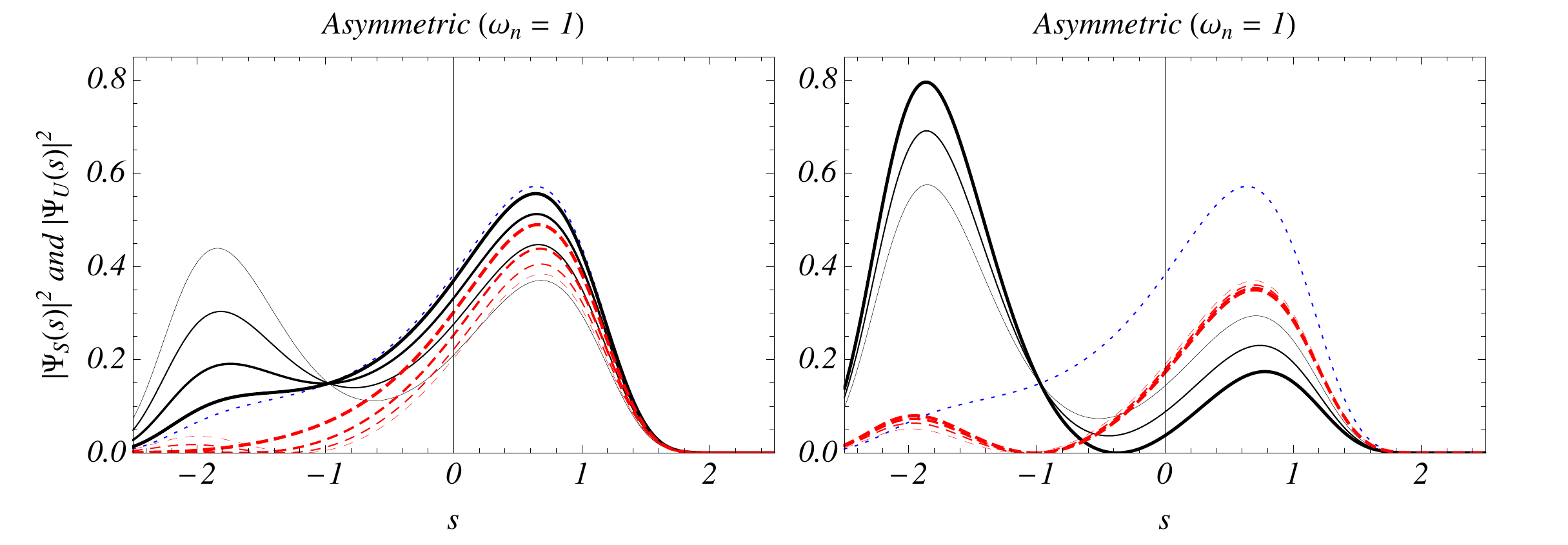}
\vspace{-.3 cm}\caption{
\small (SM)(Color online) Time evolution of probability densities for stable (solid black lines) and unstable (dashed red lines) composite states, $\Psi_S$ and $\Psi_U$, in a DW asymmetric configuration.
The curves are for times running from $T = \pi/8$ (thickest line) to $T = \pi/2$ (thinest line) in the first plot, and from $T = 5\pi/8$ (thinest line) to $T = \pi$ (thickest line) in the second plot, with steps $\Delta T = \pi/8$.
Dotted blue lines correspond to $T = 0$ for both, stable and unstable tunneling.
The results are for $\gamma = \sigma = 1$.}
\label{Wig03}
\end{figure}

\begin{figure}[h!]
\includegraphics[scale=.47]{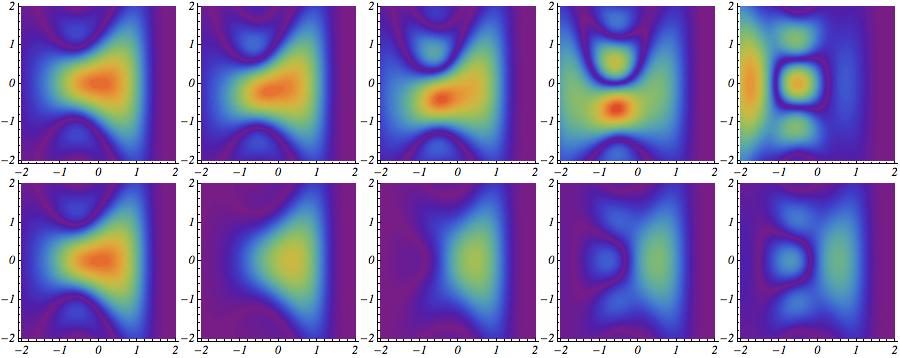}
\caption{
\small (SM)(Color online) Stable (non-stationary) and unstable (non-unitary) asymmetric tunneling evolution described by the Wigner functions (in modulus) for $\Psi_S$ (first row) and $\Psi_U$ (second row), respectively.
The plots are for $T = 0,\, \pi/8,\,\pi/4,\,\pi/2$, and $\pi$.
The contour plot follows a Rainbow color scheme from red (light gray) which corresponds to $1$, to violet (dark gray) which corresponds to $0$.
At time $T=0$ the Wigner function is assumed to be centered at the origin $(0, 0)$.
}\label{Wig04}
\vspace{-.3 cm}\end{figure}

Finally, the results for the asymmetric configuration in correspondence to the second plot from Fig.~\ref{Topo01} are shown in Figs.~\ref{Wig03} and \ref{Wig04}.

\section{Conclusions}

To conclude, our results establish the theoretical framework for obtaining the topological origin of QM vacuum transitions and DW tunneling dynamics through exactly integrable models.
On the topological front, our analysis can be extended in order to comprise the scenarios supported by a sine-Gordon theory where the primitive kink structure, $\phi = 4 \arctan(\exp[s]) - \pi$ engenders a similar QM tunneling dynamics.
On the front of the generalization of our method for understanding the tunnel effect in the context of symmetry breaking and mass generation \cite{Alex03}, the formalism can be implemented to discuss asymmetric configurations of tunneling and Hawking radiation of mass dimension one fermions \cite{RdR14} (for instance, as a dark matter candidate \cite{Alex01,Alex02}).
Moreover, the method proposed here may work as an alternative manner to provide additional solutions to the problem of tachyonic thick branes \cite{Ger01}, where a thick braneworld with a cosmological background is induced on the brane, with the respective field equations admitting a non-trivial solution for the warp factor and the tachyon scalar field \cite{Ger02,Bertolami}
Although the non-linear tachyonic scalar field generating the brane has the form of a kink-like configuration, the connection between zero-modes and first-excited states suggested in this paper may be helpful in circumventing the myriad of calculations necessary to accomplish novel localized solutions in the above-mentioned context.

{\em Acknowledgments This work was supported by the Brazilian Agency CNPq (grant 300809/2013-1 and grant 440446/2014-7).}

\pagebreak
\newpage


\end{document}